\documentclass[sigconf]{acmart}

\makeatletter
\def\@ACM@checkaffil{% Only warnings
    \if@ACM@instpresent\else
    \ClassWarningNoLine{\@classname}{No institution present for an affiliation}%
    \fi
    \if@ACM@citypresent\else
    \ClassWarningNoLine{\@classname}{No city present for an affiliation}%
    \fi
    \if@ACM@countrypresent\else
        \ClassWarningNoLine{\@classname}{No country present for an affiliation}%
    \fi
}
\makeatother

\usepackage{booktabs}
\usepackage{multirow}
\usepackage{colortbl}
\usepackage{kantlipsum} % dummy text
\usepackage{tabularx} % specify table width
\usepackage{paralist}

\AtBeginDocument{%
  \providecommand\BibTeX{{%
    \normalfont B\kern-0.5em{\scshape i\kern-0.25em b}\kern-0.8em\TeX}}}

\setcopyright{acmcopyright}
\copyrightyear{2023}
\acmYear{2023}
\acmConference[ICAIF-23]{4th ACM International Conference on AI in Finance}{November 27--29, 2023}{New York, NY}
\begin{document}

\title{Enhancing Asset Allocation in the Fixed Income Universe: A Synthetic Dataset Approach for Robust Portfolio Construction}

\title{Generating Realistic Synthetic Datasets for Improved Asset Allocation in the Fixed Income Universe}

\title{Improved Data Generation for Enhanced Asset Allocation: A Synthetic Dataset Approach for the Fixed Income Universe}

\author{Szymon Kubiak}
\affiliation{%
  \institution{City, University of London}
}

\author{Tillman Weyde}
\affiliation{%
  \institution{City, University of London}
}

\author{Oleksandr Galkin}
\affiliation{%
  \institution{City, University of London}
}

\author{Dan Philps}
\affiliation{%
  \institution{University of Warwick}
  %\institution{Rothko Investment Strategies}
}

\author{Ram Gopal}
\affiliation{%
  \institution{University of Warwick}
}

\renewcommand{\shortauthors}{Kubiak, et al.}

\begin{abstract}
    We present a novel process for generating synthetic datasets tailored to assess asset allocation methods and construct portfolios within the fixed income universe. Our approach begins by enhancing the CorrGAN model \cite{marti2020corrgan} to generate synthetic correlation matrices. Subsequently, we propose an Encoder-Decoder model that samples additional data conditioned on a given correlation matrix. The resulting synthetic dataset facilitates in-depth analyses of asset allocation methods across diverse asset universes. Additionally, we provide a case study that exemplifies the use of the synthetic dataset to improve portfolios constructed within a simulation-based asset allocation process.
\end{abstract}


\settopmatter{printacmref=false}

\maketitle

\section{Introduction}

When applying asset allocation methods in practice, a crucial challenge arises due to limited and biased empirical datasets. Conventional approaches to constructing portfolios solely on historical data often suffer from overfitting and neglect the potential market scenarios that could have occurred but did not \cite{lopez2019tactical}. To address this problem and enhance the performance of asset allocation methods during out-of-sample periods, the utilization of synthetic financial datasets has emerged as a promising solution.

Recent literature provides examples where synthetic correlation matrices were used to assess asset allocation methods in simulation-based frameworks \cite{lopez2016building, papenbrock2021matrix, marti2021ccorrgan}. However, limitations arise when applying these methods in more complex scenarios, especially when dealing with financial assets that exhibit diverse characteristics or when practitioners have objective functions and constraints that are more intricate than those of simple minimum variance portfolios. In our paper, we analyze the fixed income universe, comprising assets with varying volatility levels, where a frequent optimization goal is maximizing yield within specified risk constraints. To perform a comprehensive simulation-based assessment of asset allocation methods within this universe, it is necessary to not only have realistic correlation matrices but also other attributes, particularly volatilities and yield data that is coherent with the matrices. This necessitates the use of more advanced models for data sampling.

We propose a two-step process for generating a synthetic dataset tailored specifically to assess and fine-tune asset allocation within the fixed income universe. First, we employ a generative adversarial network to sample realistic correlation matrices, capturing the interdependencies among assets. Then, we generate additional characteristics of the assets conditional on the matrices, including volatilities, expected returns, and forward returns data, using a proprietary Encoder-Decoder model. We hypothesize that the synthetic dataset, with its reduced reliance on historical data, can address the limitations of empirical datasets and improve the performance of asset allocation methods during out-of-sample periods. Our analysis indicates that the generative models provide a more diverse dataset that extrapolates distributions into plausible areas and mitigates outliers, decreasing dependency on historical events with low future probability.

We also present a case study to showcase the practical application and potential benefits of the synthetic dataset. Our findings indicate that the synthetic dataset enables investors to construct more robust portfolios.

In summary, our research contributes to addressing the challenges posed by limited and biased empirical datasets in fixed income asset allocation. By introducing a novel process for generating a tailored synthetic dataset and demonstrating its benefits through a case study, we provide valuable insights for practitioners and researchers in the field.

\section{Related Work}

Asset allocation methods aim to distribute funds across different investments to optimize return and manage risk. Lopez de Prado introduced two asset allocation methods, Hierarchical Risk Parity \cite{lopez2016building} and Nested Clustered Optimization \cite{lopez2016robust}, and to indicate their superior performance over more traditional frameworks used a Monte Carlo approach based on synthetic correlation and associated covariance matrices. The matrices were formed using multivariate normal distributions with additional random shocks in the former paper and imposed correlated blocks in the latter article. While the performed tests provided convincing arguments supporting the introduced techniques, the synthetic data was drawn from relatively simple processes and may not be suited for more advanced simulations and tests. 

A more sophisticated method of sampling synthetic correlation matrices was proposed by Papenbrock et al. \cite{papenbrock2021matrix} who employed a multi-objective evolutionary algorithm. The generated dataset was used to compare the performance of asset allocation methods in a simulation-based framework. The authors also applied SHAP framework \cite{lundberg2017unified} to find which characteristics of correlation matrices could provide intuitive explanations of what asset allocation method to use. The approach provided realistic synthetic correlation matrices that exhibited many characteristics of empirically observed ones but the algorithm was designed to optimize for the characteristics, and as a result, it might have omitted features that were not specifically included in the objective functions.

Generative Adversarial Networks (GANs) \cite{goodfellow2020generative} have gained popularity in the financial domain for generating synthetic datasets. GANs consist of a generator and a discriminator network, where the generator generates new data samples resembling the training data, and the discriminator distinguishes between real and fake samples. GANs have been applied in various finance-related studies, such as generating time-series data for trading strategy calibration \cite{koshiyama2021generative}, addressing imbalanced classes in actuarial datasets \cite{ngwenduna2021alleviating}, and hedging \cite{kim2021deep}.

Marti \cite{marti2020corrgan} introduced a GAN model based on the DCGAN architecture \cite{radford2015unsupervised} to generate synthetic correlation matrices. \cite{marti2021ccorrgan} extended the model to condition sampled data on market regimes and conducted experiments to assess distributions of outcomes for selected asset allocation strategies. The framework learns a number of characteristics of the empirical correlation matrices in an unsupervised manner and provides high-quality data samples. The generated matrices are visually undiscernible to empirical matrices which was supported by results of an online poll in which users could guess whether depicted correlation matrices are real or fake.

\cite{mariani2019pagan} and \cite{lu2022autoencoding} proposed alternative approaches to generating synthetic data for portfolio construction based on GAN models that learn a joint probability distribution of assets' future price trends based on historical time series. The authors indicate that portfolios constructed using simulations derived from the GANs outperform traditional asset allocation methods. While the approach of optimizing portfolios across time-series simulations is promising and may potentially provide a more comprehensive environment to assess asset allocation methods than using correlation matrices, the papers provided examples with a low number of assets. Also, given that the testing of correlations across the simulations is limited, it is not clear whether the independency structure of the synthetic data exhibits all of the characteristics found in empirical datasets and listed in \cite{marti2021ccorrgan}.

Synthetic data from generative models can also be valuable in assessing portfolio risk, simulating extreme events, and testing the robustness of models to different scenarios. Flaig and Junike \cite{flaig2022scenario} employed GAN architecture to create an economic scenario generator for calculating market risk in insurance companies. In a separate publication \cite{flaig2023validation}, the authors presented methods for further validating their model, assessing the alignment of dependencies between risk factors, and measuring model overfitting to empirical data. \cite{cont2022tail} introduced the Tail-GAN model, focusing on generating tail risk scenarios for user-specified trading strategies. The model utilized a bespoke loss function to accurately capture the tail risk of benchmark portfolios based on common risk measures such as Value-at-Risk and Expected Shortfall.

In the later sections of our paper, we present a case study that utilizes the concept of tracking error volatility (TEV) which measures the standard deviation of a portfolio's returns relative to its benchmark \cite{roll1992mean}. Given the diverse characteristics of fixed income assets, investors can construct portfolios that are significantly different from their benchmarks. TEV serves as a simple measure of relative risk, particularly relevant to portfolio managers who have a mandate to outperform a benchmark within strict risk limits.

%In the later sections of our paper, we present a case study that utilizes the concept of tracking %error volatility (TEV) which measures the standard deviation of a portfolio's returns relative to %its benchmark \cite{roll1992mean}. Given the diverse characteristics of assets in the fixed %income universe, investors can construct portfolios that are significantly different from their %benchmarks. TEV serves as a simple measure of relative risk, particularly relevant to portfolio %managers who have a mandate to outperform a benchmark within strict risk limits.

\section{Generating synthetic data}
We propose a process to generate synthetic data that consists of two parts:

\begin{compactenum}
\item A generative adversarial network (GAN) model that samples synthetic correlation matrices.
\item An Encoder-Decoder model that provides additional data vectors (asset volatilities, expected returns, and forward realized returns) conditionally on a given correlation matrix.
\end{compactenum}
\vspace{\baselineskip}

The GAN model in 1) is supposed to provide information on market structure and current market conditions. The Encoder-Decoder network in 2) uses correlation matrices as input data and generates any additional information that may be required.

\begin{table*}[ht]
	\centering
	\resizebox{\linewidth}{!}{
    \begin{tabularx}{\linewidth}{l*{13}{>{\centering\arraybackslash}X}}
    \hline
     & \multicolumn{2}{c}{mean correl} & \multicolumn{2}{c}{eigen gini} & \multicolumn{2}{c}{\begin{tabular}[c]{@{}c@{}}coph corr \\ single\end{tabular}} & \multicolumn{2}{c}{\begin{tabular}[c]{@{}c@{}}coph corr \\ ward\end{tabular}} & \multicolumn{2}{c}{\begin{tabular}[c]{@{}c@{}}perron frob \\ sum neg\end{tabular}} & \multicolumn{2}{c}{\begin{tabular}[c]{@{}c@{}}power eigen\\ values\end{tabular}} \\
     & \multicolumn{1}{c}{Mean} & \multicolumn{1}{c}{Std} & \multicolumn{1}{c}{Mean} & \multicolumn{1}{c}{Std} & \multicolumn{1}{c}{Mean} & \multicolumn{1}{c}{Std} & \multicolumn{1}{c}{Mean} & \multicolumn{1}{c}{Std} & \multicolumn{1}{c}{Mean} & \multicolumn{1}{c}{Std} & \multicolumn{1}{c}{Mean} & \multicolumn{1}{r}{Std} \\ \hline
    Empirical &  &  &  &  &  &  &  &  &  &  &  &  \\ \hline
    Training Period & 0.231 & 0.063 & 0.855 & 0.011 & 0.716 & 0.118 & 0.751 & 0.071 & 2.335 & 0.389 & 1.952 & 0.093 \\
    Testing Period & 0.284 & 0.079 & 0.862 & 0.019 & 0.717 & 0.102 & 0.697 & 0.079 & 2.227 & 0.396 & 1.965 & 0.112 \\ \hline
    Generated &  &  &  &  &  &  &  &  &  &  &  &  \\ \hline
    DCGAN & 0.222 & 0.038 & 1.044 & \textbf{0.031} & 0.670 & 0.098 & 0.728 & 0.077 & 3.306 & 0.288 & 2.154 & 0.068 \\
    WGAN & \textbf{0.230} & \textbf{0.070} & \textbf{0.976} & 0.064 & \textbf{0.697} & \textbf{0.115} & \textbf{0.742} & \textbf{0.073} & \textbf{3.225} & \textbf{0.303} & \textbf{2.032} & \textbf{0.081} \\ \hline
    \end{tabularx}
}
\caption{Summary statistics of empirical and generated correlation matrices. The highlighted numbers correspond to values that are closer to the characteristics of the empirical dataset in the training period. The WGAN model provided correlation matrices that more closely resemble the empirical dataset compared to the DCGAN model across the analyzed metrics.}
\end{table*}

Our analysis is based on a time-series dataset spanning from April 2007 to October 2022. This dataset includes 68 financial assets, representing 28 currencies and 40 investable fixed income indices. The fixed income indices cover various types of treasury, corporate, government-related, and securitized bonds in developed countries across North America, Europe, Asia, as well as emerging market countries. The selection of the financial instruments aims to reflect a broad spectrum of building blocks, suitable for constructing a well-diversified global fixed income investment product. Although all selected fixed income indices are FX-hedged to USD, our case study allows to open FX positions by taking long or short positions directly on the currencies. For each asset, we collected two main time-series from Bloomberg terminal – a total return index and a series of expected returns based on assets' yields, adjusted to account for the yield that an investor could have obtained by hedging FX risk to USD with forward contracts. The generative models were trained on a transformed form of the dataset that consists of the following four components for each day in the analyzed period: correlation matrix based on 52 weekly returns, vector of volatilities based on 52 weekly returns, vector of 1-year expected returns and vector of 1-month forward realized returns.

%\begin{compactitem}
%\item correlation matrix based on 52 weekly returns
%\item vector of volatilities based on 52 weekly returns
%\item vector of 1-year expected returns
%\item vector of 1-month forward realized returns
%\end{compactitem}
%\vspace{\baselineskip}

The dataset is divided into a training period from April 2007 to March 2017 and a testing period from May 2017 to October 2022. The testing period covered a significant drawdown that the fixed income universe experienced in 2022 when both, interest rates increased and credit spreads widened. The rare combination of positively correlated interest rates and spreads accompanied by high volatility provides an interesting testing environment for any asset allocation strategy seeking to minimize risk. 

\subsection{Generative Adversarial Network}
The GAN model developed for the analysis takes a random vector and outputs a correlation matrix that exhibits characteristics found in empirically observed correlation matrices. The architecture of the model follows the methodology described in \cite{marti2020corrgan} but introduces several enhancements:

\begin{compactitem}
\item The model is converted to a Wasserstein GAN model following an approach described in \cite{gulrajani2017improved}. Without the change, the network in some configurations of hyperparameters failed to converge or exhibited mode collapse \cite{goodfellow2016nips}. The most common reason for the issues was that the discriminator was learning too quickly which did not allow the generator to improve. The change has also improved the characteristics of generated correlation matrices across the tests that were conducted.
\item In many hyperparameter configurations, the generated correlation matrices exhibited a “checkerboard artifact” described in \cite{odena2016deconvolution} and \cite{aitken2017checkerboard}. The issue may occur when there is an overlap between filters of transposed convolutional layers and consequently, some pixels (or correlation coefficients in this case) receive less focus in the learning process. To address this issue, in the generator network, we reduced the use of transposed convolutional layers and instead favored convolutional and upsampling layers with nearest-neighbor interpolation. 
\item The empirical correlation matrices used for training were not sorted using a hierarchical clustering algorithm; instead, they maintained a consistent asset order. While \cite{marti2020corrgan} suggested sorting the matrices to emphasize the hierarchical structure of financial markets, in our experiments we noticed that the GAN model can learn characteristics of the empirical dataset without reordering the empirical dataset.
\end{compactitem}
\vspace{\baselineskip}

We evaluate the GAN model by comparing the characteristics of the generated correlation matrices with empirical ones. Two variants of the GAN model are assessed: "DCGAN," the original model based on \cite{marti2020corrgan}, and "WGAN," the enhanced model proposed in this section. We sampled 64,000 correlation matrices for both models and used metrics suggested in \cite{papenbrock2021matrix} for assessment:

\begin{compactitem}
\item mean correl – average correlation coefficient across matrices
\item eigen gini – Gini coefficient of matrices’ eigenvalues
\item coph corr – cophenetic correlation between the original correlation distance matrix and the cophenetic matrix of the hierarchical clustering algorithm (using single or ward linkage). This is a proxy that measures how hierarchical correlation matrices are. The correlation distance is measured as \(D = \sqrt{2(1-Corr)}\)
\item perron frob sum neg – measures the sum of negative entries of the first eigenvector.
\item power eigen values – power exponent of correlation matrices’ eigenvalue distribution
\end{compactitem} 

\vspace{\baselineskip}

As shown in Table 1, most WGAN metrics closely resemble those of empirical matrices. Notable distinctions emerge mainly in the "perron frob sum neg" values, which signify a higher prevalence of negative elements within the eigenvector corresponding to the largest eigenvalue. Additionally, Gini coefficients and power exponents of eigenvalues exhibit higher averages compared to the empirical dataset. All WGAN variant's mean metrics are closer to those of the empirical training dataset compared to the DCGAN model. While some of the differences between models are not large, it is worth noting that training the DCGAN model necessitated hyperparameter calibration and multiple training trials to achieve successful convergence and avoid the mode collapse issue.

\begin{figure}[b]
  \centering
  \includegraphics[scale=0.60]{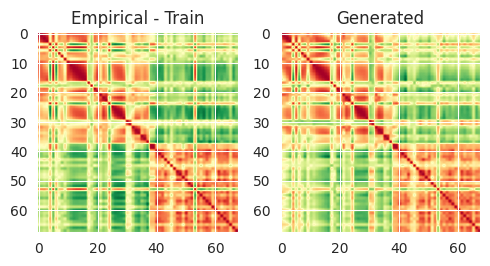}
  \caption{Sample empirical and "WGAN"-generated correlation matrices. The generated matrices' are visually indistinguishable from empirically observed correlation matrices.}
\end{figure}

The higher performance of the WGAN model can be also illustrated by comparing the diagonals of the matrices - the average diagonal element of WGAN’s matrices is 0.9992 whereas for DCGAN the average is 0.9719. While the original CorrGAN model \cite{marti2020corrgan} was trained on a different dataset, the author noted an average diagonal element of 0.998.

Similarly to \cite{marti2020corrgan}, our generated correlation matrices are difficult to be visually discerned from empirical correlation matrices as illustrated in Figure 1.

\subsection{Encoder-Decoder network for additional asset characteristics}

The Encoder-Decoder model generates additional asset attributes from a correlation matrix, including volatilities, expected returns, and 1-month forward returns. The additional attributes should reflect the market conditions that a correlation matrix corresponds to.

The encoder takes a correlation matrix as input and deconstructs it using a block of 2D convolutional layers with LeakyReLU activation functions and dropout layers. The output is then compressed using linear layers. The decoder consists of three parallel blocks, each comprising linear, 1D convolutional, and upsampling layers with LeakyReLU activation functions. The model outputs three vectors representing volatilities, expected returns, and 1-month forward returns of the underlying assets. Notably, the values in these generated vectors maintain the same order of assets as the input correlation matrix.

\begin{figure}[b]
  \centering
  \includegraphics[scale=0.78]{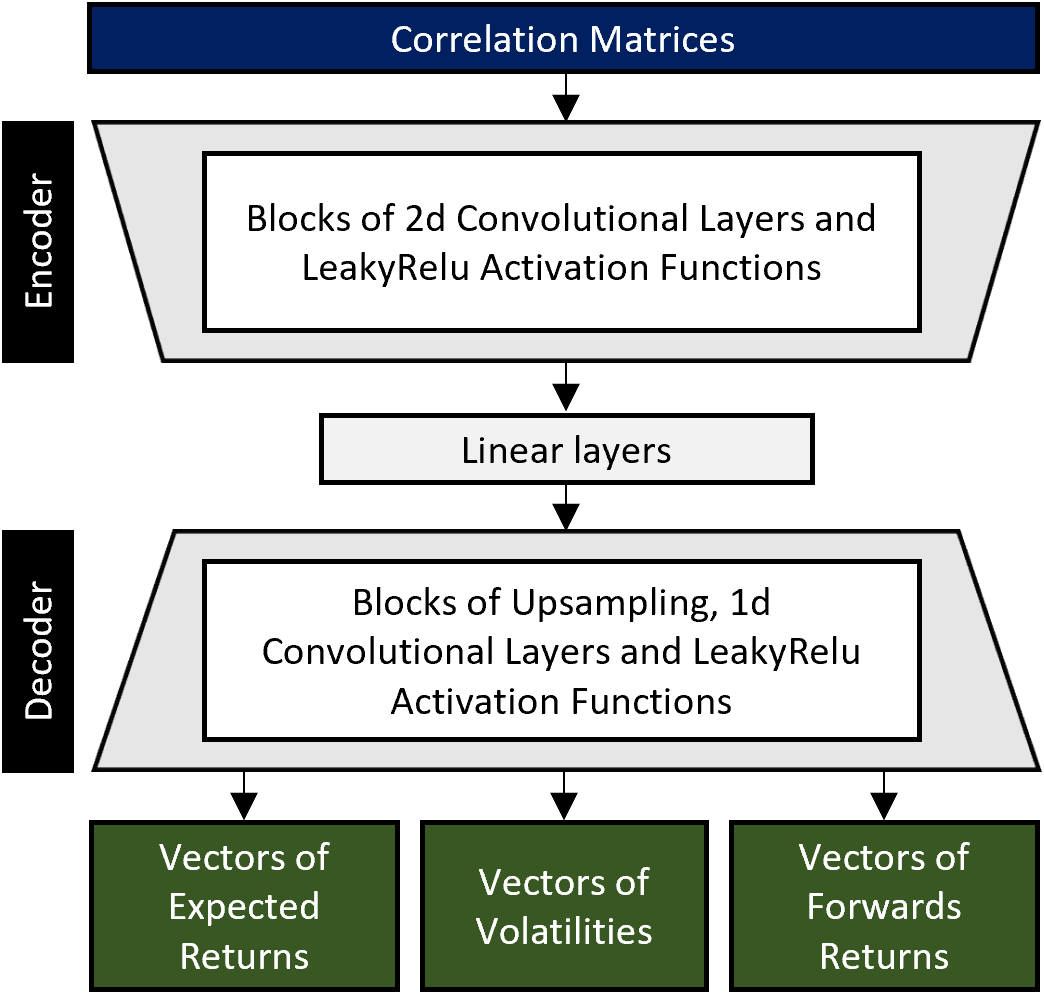}
  \caption{Scheme of the Encoder-Decoder network. The model takes a correlation matrix as input and returns three vectors that are conditional on the correlation matrix: expected returns, volatilities, and forward returns.}
\end{figure}

\begin{table}[b]
\begin{tabular}{@{}lcccc@{}}
\toprule
 & Mean & Std Dev & Skew & Kurtosis \\ \midrule
Volatilities &  &  &  &  \\ \midrule
Empirical Train & 7.098\% & 5.475\% & 1.447 & 2.807 \\
Generated & 7.391\% & 5.482\% & 1.246 & 2.048 \\
Empirical Test & 6.309\% & 5.278\% & 2.873 & 19.890 \\ \midrule
1-Year   Expected Returns &  &  &  &  \\ \midrule
Empirical Train & 3.053\% & 3.297\% & 4.613 & 96.076 \\
Generated & 2.937\% & 3.000\% & 3.073 & 35.917 \\
Empirical Test & 2.397\% & 3.828\% & 6.857 & 107.077 \\ \midrule
4-Week   Forward Returns &  &  &  &  \\ \midrule
Empirical Train & 0.251\% & 2.537\% & -0.758 & 10.601 \\
Generated & 0.223\% & 2.372\% & 0.053 & 6.026 \\
Empirical Test & -0.029\% & 2.497\% & 0.693 & 59.745 \\ \bottomrule
\end{tabular}
\caption{Summary statistics of the dataset generated by the Encoder-Decoder network. The metrics aggregate results for all assets. The characteristics of the generated dataset are similar to the empirical dataset in training period.}
\end{table}

During training, the model employs a loss function that consists of three Mean Squared Error (MSE) components to compare the generated three vectors with the original empirical dataset. Training is conducted using only empirical correlation matrices and the final model is evaluated by comparing distributions of the synthetic and empirical additional vectors. 

%\begin{equation}
%  \begin{aligned}
%    Loss = Volatilities Loss \; + \; Expected Return Loss \; + \; \\ Forward Returns Loss
%  \end{aligned}
%\end{equation}

The model acts as a deterministic "translator" that converts a correlation matrix into additional data vectors, without introducing additional randomness. The diversity in the output stems from the variety of generated correlation matrices.

For the further parts of the analysis, we created a “synthetic” dataset by sampling 64,000 correlation matrices from the GAN model and corresponding vectors of volatilities, expected returns and forward returns from the Encoder-Decoder model.

Table 2 compares the main statistics of the vectors in the synthetic and empirical datasets. The Encoder-Decoder model effectively captures the distributions of the additional data vectors, with means similar to the empirical vectors used in training. The other three moments are moderately less pronounced which is caused by the model mitigating outliers that are present in the empirical dataset. The kurtosis of forward returns is lower in the generated dataset, but still reflects fat tails in the underlying assets' returns. The generated vectors are relatively different from the empirical dataset in the testing period. However, considering that the Encoder-Decoder model did not directly target these metrics in the training process, such discrepancies were expected.

In Figure 3, we compare the distributions of the empirical and synthetic vectors for a selected asset, US High Yield, which is one of the most volatile fixed income indexes in the dataset. Some discrepancies are observed between the two datasets - notably, the Encoder-Decoder model smooths out rough sections in the empirical dataset's distributions, leading to more monotonic shapes.

\begin{figure}[ht]
  \centering
  \includegraphics[width=\linewidth]{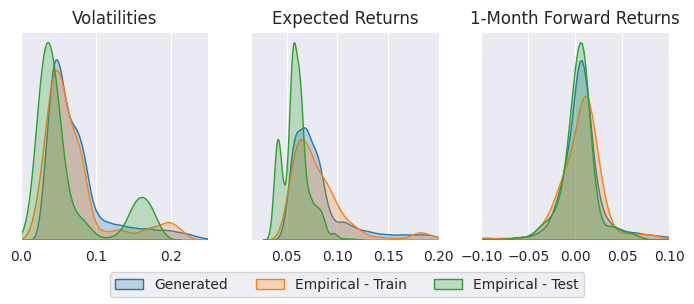}
  \caption{Distributions of additional generated vectors for US High Yield in empirical and generated datasets. The model occasionally generates "smoother" distributions than those observed in the empirical training dataset.}
\end{figure}
\begin{figure}[ht]
  \centering
  \includegraphics[width=\linewidth]{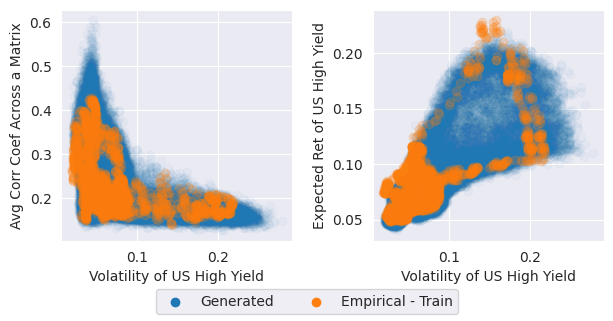}
  \caption{Scatterplots illustrating sample relationships between dataset's components for US High Yield. The Encoder-Decoder model often mitigates outliers and extrapolates into probable areas that were sparsely represented in the empirical dataset.}
\end{figure}
\begin{figure}[ht]
  \centering
  \includegraphics[width=\linewidth]{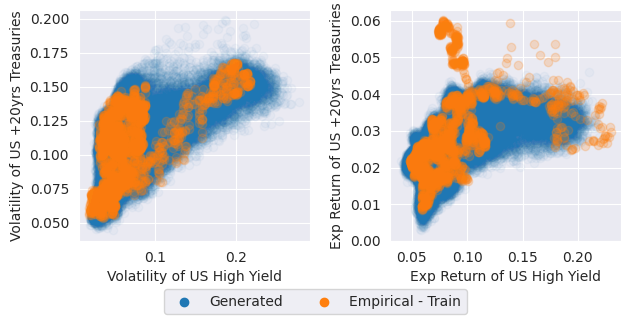}
  \caption{Scatterplots showing sample relationships between components of US High Yield and long-term US Treasuries. The components generated by the Encoder-Decoder model are conditional on a given correlation matrix and their values are consistent across different assets.}
\end{figure}

Figures 4 and 5 demonstrate that the Encoder-Decoder network learns the relationships across components for a single asset and dependencies between different assets' components. The model shows the ability to extrapolate relationships, as evident in the second scatterplot of Figure 4, where expected returns and volatilities extend into a less represented region. Although this area was not extensively covered in the empirical data, it is plausible that the asset may exhibit volatilities and expected returns within that region in the future. Overall, a visual assessment of the components across the entire dataset reveals consistent conditioning of the additional vectors by the correlation matrices for all assets.

\section{Case Study}
%\subsection{}

Let's consider an investor who aims to build a portfolio that outperforms a chosen benchmark over the long-term, while maintaining low tracking error volatility (TEV; measured as the standard deviation of the portfolio's returns relative to the benchmark). Once the initial investment is made, the portfolio allocations are not adjusted. To construct the portfolio, the study employs an approach that searches for optimal asset weights across simulations using either empirical or synthetic datasets. We hypothesize that portfolios constructed using the synthetic dataset would outperform portfolios based only on the empirical dataset in an out-of-sample period.

\begin{table*}[h]
	\centering
	\resizebox{1\linewidth}{!}{
	\begin{tabularx}{\linewidth}{l*{10}{>{\centering\arraybackslash}X}}
    \toprule
     & \#Obs & \begin{tabular}[c]{@{}c@{}}TEV\\ (bps)\end{tabular} & \begin{tabular}[c]{@{}c@{}}Excess ER\\ (bps)\end{tabular} & \begin{tabular}[c]{@{}c@{}}Average\\ Excess ER /\\ TEV Ratio\end{tabular} & t-stat & p-value & \begin{tabular}[c]{@{}c@{}}Min\\ Excess ER /\\ TEV Ratio\end{tabular} & \begin{tabular}[c]{@{}c@{}}Median\\ Excess ER /\\ TEV Ratio\end{tabular} & \begin{tabular}[c]{@{}c@{}}Max\\ Excess ER /\\ TEV Ratio\end{tabular} & \begin{tabular}[c]{@{}c@{}}Share of\\ Excess ER /\\ TEV \textgreater \\Empirical\end{tabular} \\
    \multicolumn{1}{c}{} & 1 & 2 & 3 & 4 & 5 & 6 & 7 & 8 & 9 & 10 \\ \midrule
    FR Sims &  &  & \textbf{} &  &  &  &  &  &  &  \\ \midrule
    Empirical & 335 & 74.6 & 46.6 & 0.684 &  &  & 0.106 & 0.713 & 0.943 &  \\
    Synthetic & 335 & 69.3 & 47.7 & 0.751 & -13.77 & \textless 0.01\% & \textbf{0.214} & 0.779 & 1.028 & 81.2\% \\
    Combined & 335 & \textbf{68.9} & \textbf{47.7} & \textbf{0.758} & \textbf{-14.85} & \textbf{\textless 0.01\%} & 0.202 & \textbf{0.784} & \textbf{1.050} & \textbf{84.5\%} \\ \midrule
    MV Sims &  &  &  &  &  &  &  &  &  &  \\ \midrule
    Empirical & 335 & 76.2 & 44.5 & 0.640 & \textbf{} &  & 0.096 & \textbf{0.683} & 0.990 & \textbf{} \\
    Synthetic & 335 & \textbf{74.9} & \textbf{46.4} & \textbf{0.661} & -9.52 & \textless 0.01\% & \textbf{0.203} & 0.680 & \textbf{1.036} & 73.1\% \\
    Combined & 335 & 75.0 & 46.3 & 0.661 & \textbf{-9.54} & \textbf{\textless 0.01\%} & 0.190 & 0.680 & 1.035 & \textbf{73.1\%} \\ \bottomrule
	\end{tabularx}
}
\caption{Summary statistics of the case study. The table consolidates the results from 335 experiments, comparing portfolio construction using the empirically observed ("Empirical"), synthetic ("Synthetic"), or combined ("Combined") datasets. The table is divided into two sections: FR Sims and MV Sims, representing different sources of forward returns used in simulations. The metrics examined include (2) TEV (tracking error volatility), (3) Excess ER (average expected returns in excess of a benchmark), t-statistics (5), and p-values (6), which determine whether the Excess ER to TEV ratio (4) is higher for the "Synthetic" or "Combined" variants compared to the "Empirical" variant. Additionally, the table presents the minimum, median, and maximum values of the Excess ER to TEV ratio across experiments (7, 8, 9), along with the share of observations in which the "Synthetic" and "Combined" variants outperformed the "Empirical" variants in terms of the Excess ER to TEV ratio (10).}
\end{table*}

The simulation-based asset allocation method can be described with the following process. We have an \(n x m\) matrix \(\mathbf{R}_{\text{assets}}\) that represents n simulation returns for each of the m assets. \(\mathbf{r}_{\text{bench}}\) corresponds to a returns vector of the benchmark, which is one of the columns in the matrix, and \(\mathbf{r}_{\text{port}}\) is the returns vector of the portfolio, which is obtained by multiplying the matrix of assets' simulation returns (\(\mathbf{R}_{\text{assets}}\)) by \(\mathbf{w}\), the vector of portfolio weights: $ \mathbf{r}_{\text{port}} = \mathbf{R}_{\text{assets}} \mathbf{w} $. The optimization problem can be defined as:
%\begin{equation}
%  \begin{aligned}
%        \underset{\mathbf{w}}{\text{minimize}} \left( \frac{1}{n} \sum_{i=1}^{n} \left\| \mathbf{r}_{\text{port}} - %\mathbf{r}_{\text{bench}} \right\|_2^2 \right)
%  \end{aligned}
%\end{equation}
\begin{equation}
  \begin{aligned}
        \underset{\mathbf{w}}{\text{minimize}} \left\| \mathbf{r}_{\text{port}} - \mathbf{r}_{\text{bench}} \right\|_{2}
  \end{aligned}
\end{equation}
subject to: 
%$ \mu_{\text{port}} \geq \mu_{\text{bench}} + \mathbf{t} $, $ \forall i \in M_f, \quad 0 \leq w_i \leq 1 $ and $ %\forall i \in M_c, \quad -0.05 \leq w_i \leq 0.05 $
%\begin{align*}
$\mu_{\text{port}} \geq \mu_{\text{bench}} + \mathbf{t}$,\\
%$\sum_{i=1}^{\mathbf{m}_{\text{f}}}w_{i} \leq 1$, \\
%$\forall i \in [0, m_{f}]: \quad 0 \leq w_i \leq 1$, and \\
%$\forall i \in [m_{f}+1, m]: \quad -0.05 \leq w_i \leq 0.05$ \\
%\end{align*}
%, and 
%
%$ \mathbf{w}_{\text{f}} $ \forall i \in 
where \(\mu_{\text{port}}\) and \(\mu_{\text{bench}}\) represent the average expected returns of the portfolio and benchmark across simulations and $t$ is an additional excess return target. As a result, the optimizer aims to minimize the 1-month deviations of the portfolio from the benchmark, reducing the tracking error volatility, while ensuring that the expected return is on average higher than or equal to a specified target above benchmark's expected return. The portfolio must be also fully invested, and there are limitations on FX positions within the range of -5\% to 5\%.

Simulated returns (\(\mathbf{R}_{\text{assets}}\)) and expected returns are derived from three different datasets: the empirical dataset ("Empirical" variant), the synthetic dataset ("Synthetic" variant), or a combination of both ("Combined" variant). The simulated returns were also drawn from two different methods:
\begin{compactitem}
\item "FR Sims” – the 1-month forward returns vectors were directly sourced from the empirical dataset or generated using the Encoder-Decoder model discussed earlier. This yielded 2,610 forward return sets from the empirical dataset and 64,000 from the synthetic dataset
\item “MV Sims” – returns were generated by drawing from a multivariate normal distribution, with covariance matrices based on correlation matrices and volatility vectors from either the empirical or generated datasets. In this approach, multiple simulations were performed per covariance matrix, resulting in 26,100 simulations using the empirical dataset and 640,000 simulations using the synthetic dataset
\end{compactitem}
\vspace{\baselineskip}

The "FR Sims" option was expected to more effectively incorporate the skewed distributions of assets' returns, resulting in portfolios with on average lower TEVs.

To determine if portfolios constructed using simulations based on synthetic datasets outperform, we conducted a series of experiments. These involved constructing portfolios with the described method and evaluating their performance through backtests conducted over the testing period from May 2017 to October 2022. The backtests were performed for each of the 40 fixed income indices analyzed, with varying excess expected return targets ranging from 20 to 100 bps in 10 bps increments. Instances, where the optimizer failed to converge, were excluded from the analysis. This typically occurred when the portfolio's benchmark was a high-yielding asset and there were insufficient other assets in the dataset with expected returns meeting the excess return target. Consequently, in total, 335 examples were used in the analysis in both the "FR Sims" and "MV Sims" configurations.

The primary metric used to assess the results was the ratio of expected returns in excess of the benchmark's returns to tracking error volatility. This metric directly aligns with the objectives targeted by the optimizer and serves as an indicator of the expected information ratio in the long term.

Table 4 presents the results of the case study. The variants utilizing the synthetic dataset ("Synthetic" and "Combined") demonstrated lower TEVs and higher average excess expected returns compared to the "Empirical" variant. This difference was particularly notable in the "FR Sims" options, where the TEVs of the "Synthetic" and "Combined" variants were 5 bps lower than that of the "Empirical" variant. The combination of lower TEVs and slightly higher excess expected returns led to a significant increase in the excess expected returns to TEV ratios. In the "MV Sims" options, the differences were more subtle, with the TEV being only about 1 bps lower and excess expected returns approximately 1 bps higher for the "Synthetic" and "Combined" variants compared to the "Empirical" variant. The improvements in the excess expected returns to TEV ratios were confirmed by one-sided t-tests, which yielded p-values lower than 1\%. Notably, the "Combined" variants exhibited higher t-stats than the "Synthetic" variants in both the "FR Sims" and "MV Sims" options, suggesting that augmenting the empirical dataset may offer more benefits to practitioners than solely replacing it with a synthetic dataset. 

%Figure 7 illustrates probability density functions of excess expected returns to TEV ratios %across experiments. In the "FR Sims" option, the differences between variants are evident. In the %"MV Sims" option, the "Combined" and "Synthetic" distributions appear visually identical, with %only slight improvements over the "Empirical" variant observed on the left side of the %distribution.

%\begin{figure}[ht]
%  \centering
%  \includegraphics[width=\linewidth]{ERtoTEV_PDF.png}
%  \caption{Probability density functions of excess expected returns to TEV ratios for experiments %conducted in the case study 2. In the "FR Sims" option, the distribution of the "Empirical" %variant significantly differs from the distributions of the "Synthetic" and "Combined" variants. %In the "MV Sims" option, the "Empirical" variant is slightly shifted to the left.}
%\end{figure}

\section{Conclusion}
In this work, we proposed a process to generate synthetic data for analyses of asset allocation methods in the fixed income universe. The data generation process was divided into two stages: firstly, we generated correlation matrices with an enhanced CorrGAN \cite{marti2020corrgan} model and then we sampled additional assets attributes (volatilities, expected returns and forward returns) with an Encoder-Decoder model introduced in this paper. The generated dataset exhibited characteristics of the empirical dataset and preserved relationships between the generated matrices and the additional attributes. Importantly, the generative models also demonstrated their capability in mitigating outliers, reducing reliance on historical events with low future probability, and extrapolating data into probable areas that were sparsely represented in the empirical dataset. In the case study we also illustrated that utilizing the synthetic dataset led to the construction of better-performing portfolios.

Future research endeavors could explore additional modifications to the GAN and Encoder-Decoder model, such as conditioning them on external factors to improve the performance of asset allocation methods in specific market regimes. Additionally, further investigations can focus on expanding the application of synthetic datasets to other asset classes and considering different practitioner requirements. For example, in an equities universe, the model could incorporate valuation metrics alongside volatility metrics to evaluate asset allocation methods with a value-bias.

By introducing this approach to synthetic data generation and demonstrating its practical benefits, we aim to encourage practitioners and researchers to embrace synthetic datasets as a valuable tool for improving asset allocation in the fixed income universe, as well as exploring their potential applications addressing diverse requirements of asset allocators.

\bibliographystyle{ACM-Reference-Format}
\bibliography{sample-base}

\end{document}